\documentclass[pra, twocolumn]{revtex4}

\usepackage{amsmath,amssymb,amscd}  
\usepackage{psfrag}  
\usepackage[dvips]{epsfig}  
\usepackage{epsfig}  

\def\figurelib{.}

\usepackage{enumerate}
\usepackage{color}

\usepackage{theorem}
\theorembodyfont{\rmfamily}

\theoremstyle{break}

\def\QED{~\rule[-1pt]{5pt}{5pt}\par\medskip}

\def\R{\mathbb{R}}

\def\su{\mathfrak{su}}

\def\one{\mathbf{1}}
\def\Eq{Eq.~\eqref}
\def\Eqs{Eqs.~\eqref}

\DeclareMathOperator{\tr}{Tr}

\DeclareMathOperator{\diag}{diag}

\newcommand{\ma}[1]{\left[\begin{matrix} #1 \end{matrix}\right]}

\def\ie{{\it i.e.}} 
\def\eg{{\it e.g.}} 

\begin{document}   
\title{Robust control pulses design for electron shuttling in solid
  state devices}

\author{Jun Zhang$^{1,2}$, Loren Greenman$^2$, Xiaotian Deng$^2$, and
  K. Birgitta Whaley$^2$} \affiliation{$^1$Joint Institute of
  UMich-SJTU, Shanghai Jiao Tong University, and Key Laboratory of
  System Control and Information Processing, Ministry of Education,
  Shanghai, 200240,
  China\\
  $^2$Department of Chemistry, Berkeley Center for Quantum Information
  and Computation, University of California, Berkeley, California
  94720, USA}

\date{October 10, 2012}

 \begin{abstract}  
   In this paper we study robust pulse design for electron shuttling in
   solid state devices. This is crucial for many practical
   applications of coherent quantum mechanical systems. Our objective
   is to design control pulses that can transport an electron along a
   chain of donors, and also make this process robust to parameter
   uncertainties. We formulate it as a set of optimal control problems
   on the special unitary group SU(n), and derive explicit expressions
   for the gradients of the aggregate transfer fidelity.  Numerical
   results for a donor chain of ionized phosphorus atoms in bulk
   silicon demonstrate the efficacy of our algorithm.
 \end{abstract}

\maketitle

\section{Introduction}
Recent years have witnessed the rapid advance of solid state devices
that take full advantage of coherent quantum mechanical
properties~\cite{Kane:98,Kane:03,Morton:11,Vrijen:00,Goswami:07}.  One
particular application of such devices is quantum computation, which
has attracted intensive research interest over the past 15 years.  To
implement these devices in practical applications, a central task is
to generate a quantum state transfer. For example, it is often desired
to transfer the population between different energy levels or
different spatial locations so that quantum information can be
circulated and processed on a large scale.

In this paper we consider the design of robust control pulses for
coherent electron shuttling in solid state devices with a
one-dimensional array of quantum dots or donors.  There are a number
of proposals which use the donated electron of Group V dopants such as
phosphorus in silicon as qubits~\cite{Kane:98,Kane:03,Morton:11}.
Such dopants can be inserted using ion
implantation~\cite{Schenkel:03,Dzurak:07,Dzurak:12}.  Recently
developed techniques using a scanning tunneling microscope have
further allowed them to be placed with high
precision~\cite{Dzurak:12,Simmons:03}.  In order to couple donors and
achieve greater scalability of qubit array size, it is desirable to be
able to move qubits robustly between distant physical
locations~\cite{Kane:03}.  The objective is to transport the electron
along a chain of donors so that the encoded quantum information can be
communicated between distant qubits.  Specifically, at the beginning
of this procedure, an electron is localized at one end of the chain.
Then by applying some appropriate external control fields, we seek to
shuttle the electron to the other end of the chain.  Depending on the
specific physical implementation, the control fields can be gate
voltage~\cite{Greentree:04,Rahman:10} or tunable on-site
energy~\cite{Chen:11}.

To achieve such electron shuttling, Greentree et
al.~\cite{Greentree:04} have proposed to use a solid state version of
the well-known Stimulated Raman Adiabatic Passage (STIRAP) for
population transfer in quantum optics~\cite{Vitanov:01,Bergmann:98}.
In this adaptation to the solid state, which is known as Coherent
Tunneling Adiabatic Passage (CTAP), two Gaussian pulses are applied in
a {\it counter-intuitive} sequence to realize the population transfer
in an adiabatic manner, \ie, starting in an eigenstate of the system
Hamiltonian and changing the Hamiltonian sufficiently slowly so that
the system will remain in the corresponding eigenstate during the
entire transfer process. The amplitudes, peak times, and standard
deviations of the two CTAP pulses have to be carefully tuned.  We have
recently shown that adiabaticity is, however, not a requisite
condition for achieving high fidelity electron shuttling between
spatial locations.  In Ref.~\cite{Zhang:12}, we have applied
Lie-Poisson reduction to develop a geometric control approach to
remove the adiabatic condition and to accomplish the quantum state
transfer with complete fidelity.

In the current paper we are interested in the design of control pulses
for electron shuttling that are robust with respect to relevant
experimental parameters. In many experiments it is inevitable that
some physical parameters are not precisely known although we may have
confidence that they lie in a certain range. This makes it
particularly important to design control pulses in a robust manner so
that the electron shuttling process is insensitive to these parameter
uncertainties. We formulate this here as an optimal control problem on
the special unitary Lie group. We then discretize the uncertainty
range and obtain a finite collection of state transfer problems, each
of which takes a different value of the uncertainty parameter. The
gradients of the aggregate fidelity with respect to these control
fields are then derived in an analytic form, which allows for
efficient implementations of gradient types of optimization
algorithms. We demonstrate the efficiency of our algorithm here by
numerical studies with realistic physical parameters relevant to the
electron shuttling between phosphorus dopant ions in silicon.

\section{Problem formulation}
In this section we provide a general mathematical description for
electron shuttling in solid state devices, together with the key
associated mathematical background.

The underlying physics and potential applications of solid state
devices with qubits have been widely discussed in the physical
community. See, \eg, Refs.~\cite{Greentree:04,Rahman:10,Chen:11}. For
a complete quantum description of the system under realistic
conditions, it is necessary to employ the density operator $\rho$,
which is a Hermitian matrix with unit trace.  The diagonal elements of
the density operator correspond to the electron populations on each
site.  The dynamics of the density operator is determined by the
Liouville-von Neumann equation:
\begin{equation}
  \label{eq:35}
   \dot \rho=-[iH, \rho],
\end{equation}
where $H$ is a traceless Hermitian matrix which is termed the system
Hamiltonian.  To be specific, we focus here on a triple donor system,
but note that the development and solution shown here can be easily
extended to devices with more donors. In this case, the term $iH$ is
defined on the Lie algebra $\su(3)$, \ie, all the $3\times 3$
skew-Hermitian matrices.  In Ref.~\cite{Greentree:04}, an electron is
moved between ends of a chain of ionized phosphorus dopants, for which
the Hamiltonian is given by (setting $\hbar=1$):
\begin{equation}
  \label{eq:33}
  H=\ma{0&-\Omega_{12}&0\\-\Omega_{12}&\Delta&-\Omega_{23}\\
0&-\Omega_{23}&0}.
\end{equation}
Here $\Delta$ is the energy difference between eigenstates, and
$\Omega_{12}$ and $\Omega_{23}$ are the coherent tunneling amplitudes
between eigenstates.

Define a basis for $\su(3)$ as
\begin{equation}
  \label{eq:2}
    \begin{aligned}
X_1&=\ma{0&i&0\\i&0&0\\0&0&0},  &X_2&=\ma{0&0&0\\0&0&i\\0&i&0},\\
X_3&=\ma{0&0&1\\0&0&0\\-1&0&0}, &X_4&=\ma{0&1&0\\ -1&0&0\\0&0&0}, \\
X_5&=\ma{0&0&0\\0&0&1\\0&-1&0}, &X_6&=\ma{0&0&i\\0&0&0\\i&0&0}, \\
X_7&=\ma{i&0&0\\0&-i&0\\0&0&0}, &X_8&=\frac{1}{\sqrt{3}}\ma{i&0&0\\0&i&0\\0&0&-2i}.
  \end{aligned}
\end{equation}
With a rearrangement of order, this choice of $\su(3)$ basis is seen
to be equivalent to the Gell-Mann matrices~\cite{Georgi:99}.  In this
basis, the Hamiltonian in \Eq{eq:33} can be represented as
\begin{equation}
  \label{eq:34}
  iH=-\Omega_{12} X_1-\Omega_{23}X_2-\frac{\Delta}2 X_7
+\frac{\Delta}{2\sqrt{3}}X_8+\frac{\Delta}3 I_3,
\end{equation}
where $I_3$ is the $3\times 3$ identity matrix.  We can drop the term
$\frac{\Delta}3 I_3$ since it commutes with all the other terms and thus
contributes only a global phase.

Without loss of generality, let us denote the spatial state of the left end
of the chain as
\begin{equation}
  \label{eq:6}
  \rho_I=\ma{1&0&0\\0&0&0\\0&0&0}
\end{equation}
and the right end of the chain as
\begin{equation}
  \label{eq:7}
  \rho_T=\ma{0&0&0\\0&0&0\\0&0&1}.
\end{equation}
The electron shuttling can now be formulated as a steering problem,
that is, for the dynamical system of \Eq{eq:35}, we will apply coherent
tunneling amplitudes $\Omega_{12}$ and $\Omega_{23}$ as control fields to
transfer the density matrix $\rho$ from the initial state $\rho_I$ at
the initial time $t=0$ to the final state $\rho_T$ at the terminal
time $t=T$.

For a fixed energy difference $\Delta$, this problem has been solved
by the same authors in~\cite{Zhang:12}. In that work we developed an
efficient numerical algorithm by using the Lie-Poisson reduction
theorem.  However, as noted above, in real experiments, it is often
the case that the exact value of $\Delta$ cannot be determined
precisely, e.g., due to imperfections in engineering implementations.
Instead, we may only know that the energy difference $\Delta$ lies in
a range $[\Delta^*-\Delta_\epsilon, \Delta^*+\Delta_\epsilon]$, where
$\Delta^*$ is the nominal value and $\Delta_\epsilon$ is the maximum
possible error bound. These two values are usually available for a
specific physical system.

In the rest of this paper, we will design robust control pulses that
can achieve the desired spatial state transfer {\it regardless} of
what the true energy difference is in the given interval.

\section{Robust optimal control algorithm}
\label{sec:oc}
To solve the aforementioned robust state transfer problem, we take a
number of sampling points in the uncertainty interval and then form a
collection of state transfer problems, each of which has a different
energy difference. We then apply a gradient algorithm to find the
optimal solution that solves all these problems simultaneously.

To this end, we take $N$ equally spaced points $\{\Delta_n\}_{n=1}^N$ in
the uncertainty interval $[\Delta^*-\Delta_\epsilon,
\Delta^*+\Delta_\epsilon]$, that is,
\begin{equation*}
\Delta_n=\Delta^*-\Delta_\epsilon+\frac{2(n-1)}{N-1} \Delta_\epsilon,
\end{equation*}
and $\Delta_1=\Delta^*-\Delta_\epsilon$,
$\Delta_N=\Delta^*+\Delta_\epsilon$. For each $\Delta_n$, we consider
a dynamical system with the Liouville-von Neumann equation
\begin{equation}
  \label{eq:4}
   \dot \rho_n=-[iH_n, \rho_n],  
\end{equation}
where the Hamiltonian $H_n$ is given by
\begin{equation}
  \label{eq:3}
  iH_n=-\Omega_{12} X_1-\Omega_{23}X_2-\frac{\Delta_n}2 X_7
+\frac{\Delta_n}{2\sqrt{3}}X_8.
\end{equation}
We now have a set of $N$ dynamical systems, which are all
identical except for a different value of $\Delta$ in each case.

We want to steer all these $N$ dynamical systems from the initial
condition $\rho_I$ in \Eq{eq:6} to the final state $\rho_T$ in
\Eq{eq:7}. Denote the state trajectory of $n$-th system as $\rho_n$.
We can formulate the state transfer for this system as the following minimization
problem:
\begin{equation}
  \label{eq:9}
 \min L_n=\|\rho_T-\rho_n(T)\|_F^2,
\end{equation}
where the Frobenius norm is defined as
\begin{equation}
  \label{eq:13}
  \|A\|_F^2=\tr AA^\dag.
\end{equation}
We then have
\begin{equation*}
  \label{eq:14}
  \begin{aligned}
L_n=&\tr (\rho_T-\rho_n(T))(\rho_T-\rho_n(T))^\dag \\
=&\tr \rho_T\rho_T^\dag +\tr \rho_n(T)\rho_n^\dag(T) \\
&\quad-\tr \rho_T \rho_n^\dag(T)-\tr \rho_n(T) \rho_T^\dag.
  \end{aligned}
\end{equation*}
It is easy to show that $\rho=\rho^\dag$ and $\tr
\rho_n(T)\rho_n^\dag(T)=1$, and thus minimizing $L_n$ amounts to
maximizing the following fidelity function
\begin{equation}
  \label{eq:15}
  \max J_n=\tr\rho_T\rho_n(T).
\end{equation}
The robust state transfer can now be formulated as maximization of the
aggregate fidelity of all the terminal states
$\rho_n(T)$:
\begin{equation}
  \label{eq:16}
\max  J=\sum_{n=1}^N J_n = \sum_{n=1}^N\tr \rho_T\rho_n(T).
\end{equation}

\subsection{Discretization of sinusoidal control fields}
As discussed earlier, we use the coherent tunneling amplitudes
$\Omega_{12}$ and $\Omega_{23}$ as control fields. In real physical
experiments, there usually exist maximum frequency limits on the
control signals. We therefore express the control fields as a finite
summation of harmonics:
\begin{equation}
  \label{eq:1}
  \begin{aligned}
\Omega_{12}(t)&=a_0+\sum_{m=1}^M [a_m\cos m\omega t+b_m\sin m\omega
t],\\
\Omega_{23}(t)&=c_0+\sum_{m=1}^M [c_m\cos m\omega t+d_m\sin m\omega t],
  \end{aligned}
\end{equation}
where $\omega=2\pi/T$. Here the expansions are truncated at a value $M$, which
can be chosen so that $M\omega$ stays within the feasible
frequency range. In the case when $M$ is sufficiently large, \Eq{eq:1}
can approximate any continuous control function.

For time varying control fields, there is generally no analytic method
to solve the Liouville-von Neumann equation \Eq{eq:4}. To obtain
numerical solutions, a common practice is to divide the total time
duration into a number of small time steps and assume that the control
functions are constant within each step. In particular, for a given
time duration $[0, T]$, divide it into $K$ equal intervals $\{[t_k,
t_{k+1}]\}_{k=0}^{K-1}$ of length $\Delta t=t_{k+1}-t_k=T/K$, where
$t_k=k \Delta t$.  On each of these intervals $[t_k, t_{k+1}]$, assume
the control fields in \Eq{eq:1} take constant values which are equal
to those on the left boundary $t=t_k$:
\begin{equation}
  \label{eq:10}
    \begin{aligned}
\Omega_{12}(k)&=a_0+\sum_{m=1}^M \left[a_m\cos mk\frac{2\pi}K+b_m\sin
  mk\frac{2\pi}K \right],\\
\Omega_{23}(k)&=c_0+\sum_{m=1}^M \left[c_m\cos mk\frac{2\pi}K+d_m\sin
  mk\frac{2\pi}K \right].
  \end{aligned}
\end{equation}
From \Eq{eq:3}, we obtain
\begin{equation}
  \label{eq:36}
  iH_n(k)=-\Omega_{12}(k) X_1-\Omega_{23}(k)X_2-\frac{\Delta_n}2 X_7
+\frac{\Delta_n}{2\sqrt{3}}X_8.
\end{equation}
Since $iH_n(k)$ is constant on the interval $[t_k, t_{k+1}]$, we can
compute its unitary propogator as
\begin{equation}
  \label{eq:17}
  U_n(k)=e^{-i H_n(k)\Delta t}.
\end{equation}
It follows that the density operator at the final time can be
calculated as
\begin{equation}
\label{eq:20}
\rho_n(T)=U_n(K-1) \cdots U_n(0) \rho_I U_n^\dag(0) \cdots U_n^\dag(K-1). 
\end{equation}
To realize the desired robust spatial state transfer, we now only need to
maximize the aggregate fidelity in \Eq{eq:16} with respect to the
expansion coefficients $a_m$, $b_m$, $c_m$, and $d_m$ in \Eq{eq:1}.

\subsection{Gradient derivations}
We want to apply a gradient algorithm to find the maximizing expansion
coefficients. To this end, we need to calculate the derivatives of the
cost function $J$ with respect to those expansion coefficients.

For the ease of notation, let
\begin{equation}
  \label{eq:11}
  \begin{aligned}
{\Omega_{12}}&=\ma{\Omega_{12}(0)& \cdots & \Omega_{12}(K-1)}^T, \\
{\Omega_{23}}&=\ma{\Omega_{23}(0)& \cdots & \Omega_{23}(K-1)}^T, \\
p&=\ma{a_0 & a_1 &\cdots&a_M&b_1&\cdots&b_M}^T,\\
q&=\ma{c_0 & c_1 &\cdots&c_M&d_1&\cdots&d_M}^T,\\
v_K&=\ma{0 & 1& \cdots &K-1}^T,\\
v_M&=\ma{1&\cdots &M}^T.
  \end{aligned}
\end{equation}
Then the control fields in \Eq{eq:10} can be rewritten in the
following vector form:
\begin{equation}
  \label{eq:8}
  \begin{aligned}
    \Omega_{12}(k)&=\ma{1& \cos \left(k v_M^T\dfrac{2\pi}K\right) &
      \sin \left(k v_M^T \dfrac{2\pi}K\right)} p,\\
   \Omega_{23}(k)&=\ma{1& \cos \left(k v_M^T\dfrac{2\pi}K\right) &
      \sin \left(k v_M^T \dfrac{2\pi}K\right)} q,\\
  \end{aligned}
\end{equation}
where the matrix functions $\cos(\cdot)$ and $\sin(\cdot)$ are
calculated element-wise.  Define
\begin{equation*}
G=\ma{\one &\cos \left(v_Kv_M^T\dfrac{2\pi}K\right) & 
\sin \left(v_Kv_M^T \dfrac{2\pi}K\right) },
\end{equation*}
where $\one$ is a column vector with all entries being $1$.  Then we
have
\begin{equation}
  \label{eq:18}
  \Omega_{12}=Gp, \quad \Omega_{23}=G q.
\end{equation}

Now the optimization variables become two vectors $p$ and $q$, both of
which lie in $\R^{M+1}$. We proceed to derive the gradients of the
aggregate fidelity $J$ with respect to $p$ and $q$. From \Eqs{eq:16}
and~\eqref{eq:18}, we have
\begin{equation*}
\frac{dJ}{dp}=\left(\frac{d\Omega_{12}}{dp}\right)^T
\frac{dJ}{d\Omega_{12}}=G^T \sum_{n=1}^N \frac{dJ_n}{d\Omega_{12}}.
\end{equation*}
Similarly,
\begin{equation*}
\frac{dJ}{dq}=G^T \sum_{n=1}^N \frac{dJ_n}{d\Omega_{23}}.
\end{equation*}
Next we need to derive $\frac{dJ_n}{d\Omega_{12}(k)}$ and
$\frac{dJ_n}{d\Omega_{23}(k)}$. Define
\begin{equation*}
  \begin{aligned}
\rho_n(k)&=U_n(k-1) \cdots U_n(0) \rho_I U_n^\dag(0) \cdots U_n^\dag(k-1),\\
\Lambda_n(k)&=U_n^\dag(k) \cdots U_n^\dag(K-1) \rho_T U_n(K-1) \cdots U_n(k),
  \end{aligned}
\end{equation*}
where $U_n(k)$ is defined in \Eq{eq:17}, and $k=0$, \dots, $K-1$. In
addition, define $\rho_n(0)=\rho_I$ and $\Lambda_n(K)=\rho_T$.  Then
\begin{equation*}
  \label{eq:37}
  \begin{aligned}
J_n&=\tr \Lambda_n(K) \rho_n(K)=\tr \Lambda_n(K-1)\rho_n(K-1) \\
&= \cdots=\tr \Lambda_n(1) \rho_n(1)=\tr \Lambda_n(0)\rho_n(0).
  \end{aligned}
\end{equation*}
It follows that
\begin{equation}
 \label{eq:38}
 \begin{aligned}
&\frac{d J_n}{d \Omega_{12}(k)}=\frac{d\tr \Lambda_n(k+1)\rho_n(k+1)}{d \Omega_{12}(k)}\\
=&\frac{d\tr \Lambda_n(k+1) U_n(k) \rho_n(k)U_n^\dag(k)}{d \Omega_{12}(k)} \nonumber\\
=&\tr \Lambda_n(k+1) \\
&\times \left(\frac{d U_n(k)}{d \Omega_{12}(k)}\rho_n(k)U_n^\dag(k)
+ U_n(k)\rho_n(k)\frac{dU_n^\dag(k)}{d \Omega_{12}(k)}\right).   
 \end{aligned}
\end{equation}
Using the following expression for the derivative of a matrix
exponential~\cite{Najfeld:95},
\begin{equation}
  \label{eq:39}
  \left.\frac{d}{dv} e^{-i(H_a+vH_b)t}\right|_{v=0}
=-i \int_0^t e^{-i H_a\tau} H_b e^{i H_a\tau} d\tau\ e^{-i H_a t},
\end{equation}
we obtain
\begin{eqnarray}
  \label{eq:40}
  \frac{d U_n(k) }{d \Omega_{12}(k)}  
= \int_0^{\Delta t} e^{-i H_n(k)\tau} X_1 e^{i H_n(k)\tau} d\tau\
U_n(k). 
\end{eqnarray}
Substituting \Eq{eq:40} into \eqref{eq:38}, we get
\begin{equation}
  \label{eq:41}
  \begin{aligned}
    &\frac{d J_n}{d \Omega_{12}(k)}\\
=&\tr \Lambda_n(k+1) \left(
\int_0^{\Delta t} e^{-i H_n(k)\tau}X_1 e^{i H_n(k)\tau} d\tau
 \rho_n(k+1)\right. \\
&\left.\qquad\qquad-\rho_n(k+1) \int_0^{\Delta t} e^{-i H_n(k)\tau} X_1 e^{i H_n(k)\tau} d\tau
 \right) \\
=& \tr[\rho_n(k+1), \Lambda_n(k+1)]\int_0^{\Delta t} 
e^{-i H_n(k)\tau} X_1 e^{i H_n(k)\tau}d\tau. \\
  \end{aligned}
\end{equation}
We can further simplify the calculation of \Eq{eq:41}. We first note that since
$H_n(k)$ is a Hermitian matrix, it can be diagonalized as
\begin{equation}
  \label{eq:22}
    H_n(k)=T_n(k) \Gamma_n(k) T_n^\dag(k),
\end{equation}
where 
\begin{equation*}
  \begin{aligned}
\Gamma_n(k)&=\diag\{\gamma_n^1(k), \gamma_n^2(k), \gamma_n^3(k)\}\\
&=\diag \left\{-\frac{\Delta_n}3, \frac{\Delta_n+3{g_n(k)}}6,
    \frac{\Delta_n-3{g_n(k)}}6 \right\},
  \end{aligned}
\end{equation*}
and the unitary matrix  $T_n(k)$ can be written as

\begin{widetext}
\begin{equation*}
  \label{eq:5}
  T_n(k)=\ma{-\Omega_{23}(k)/{h(k)} &
 \Omega_{12}(k)/\sqrt{{g_n(k)}({g_n(k)}+\Delta_n)/2}& 
\Omega_{12}(k)/\sqrt{{g_n(k)}({g_n(k)}-\Delta_n)/2}\\
0 & -\sqrt{({g_n(k)}+\Delta_n)/(2{g_n(k)})} & \sqrt{({g_n(k)}-\Delta_n)/(2{g_n(k)})} \\
\Omega_{12}(k)/{h(k)} & \Omega_{23}(k)/\sqrt{{g_n(k)} ({g_n(k)}+\Delta_n)/2}
&\Omega_{23}(k)/\sqrt{{g_n(k)}({g_n(k)}-\Delta_n)/2)}},
\end{equation*}
\end{widetext}
and 
\begin{equation*}
  \begin{aligned}
{g_n(k)}&=\sqrt{\Delta_n^2+4\Omega_{23}^2(k)+4\Omega_{12}^2(k)},\\
{h(k)}&=\sqrt{\Omega_{23}^2(k)+\Omega_{12}^2(k)}.
  \end{aligned}
\end{equation*}
Therefore we can write,
\begin{equation}
 \label{eq:21}
  \begin{aligned}
&  \int_0^{\Delta t} e^{-i H_n(k)\tau} X_1 e^{i H_n(k)\tau}d\tau\\
=& \int_0^{\Delta t} T_n(k) e^{-i \Gamma_n(k)\tau} T_n^\dag(k)X_1 T_n(k)
e^{i \Gamma_n(k)} T_n^\dag(k) d\tau \\
=&T_n(k) \int_0^{\Delta t} (T_n^\dag(k) X_1 T_n(k))\odot \Psi_n(k) d\tau\
T_n^\dag(k), 
  \end{aligned}
\end{equation}
where $\odot$ denotes the Hadamard product, \ie, element-wise product, of
two matrices, and the $ab$-th element of $\Psi_n(k)$ is 
$\exp\{i(\gamma_n^b(k)-\gamma_n^a(k))\tau\}$. 
Now define a matrix $\Phi_n(k)$,
whose $ab$-th element is given by
\begin{equation*}
  \begin{aligned}
\Phi_n^{ab}(k)=& \int_0^{\Delta t} \Psi_n^{ab}(k) d\tau \\
=&
\begin{cases}
\dfrac{\exp\{i(\gamma_n^b(k)-\gamma_n^a(k))\Delta t\}-1}
{i(\gamma_n^b(k)-\gamma_n^a(k))}, & \text{for } a\neq b. \\
\Delta t, & \text{for } a=b.
\end{cases}
  \end{aligned}
\end{equation*}
This allows \Eq{eq:21} to be calculated explicitly:
\begin{equation}
\label{eq:24}
\begin{aligned}
 &\int_0^{\Delta t} e^{-i H_n(k)\tau} X_1 e^{i H_n(k)\tau}d\tau\\
=&T_n(k) ( (T_n^\dag(k) X_1 T_n(k)) \odot \Phi_n(k) ) T_n^\dag(k),
\end{aligned}
\end{equation}
which in turn yields that
\begin{equation}
  \label{eq:27}
  \begin{aligned}
&\frac{d J_n}{d \Omega_{12}(k)}\\
=& \tr ([\rho_n(k), \Lambda_n(k)] T_n(k) 
( (T^\dag_n(k) X_1 T_n(k)) \odot \Phi ) T^\dag_n(k)).
  \end{aligned}
\end{equation}
A similar analysis leads to
\begin{equation}
  \label{eq:56}
  \begin{aligned}
& \frac{d J_n}{d \Omega_{23}(k)}\\
=& \tr ([\rho_n(k), \Lambda_n(k)] T_n(k) 
( (T^\dag_n(k) X_2 T_n(k)) \odot \Phi ) T^\dag_n(k)).
\end{aligned}
\end{equation}

We have derived closed form formulae for ${dJ}/{dp}$ and ${dJ}/{dq}$,
\ie, the gradients of the aggregate fidelity $J$ with respect to the
expansion coefficient vectors $p$ and $q$. It is now straightforward to
implement gradient types of algorithms such as the gradient descent
algorithm, the conjugate gradient algorithm, or the BFGS
algorithm~\cite{Polak:97}.

\section{Numerical optimization}
\begin{figure}[htb]
\centering
 \footnotesize
\psfrag{Site Population}[][]{Site Population} 
\psfrag{Tunneling Rate (meV)}[][]{Tunneling Rate (meV)} 
 \psfrag{Time (ns)}[][]{Time (ns)} 
  \psfrag{p11}[][]{$\rho_{11}$} 
  \psfrag{p22}[][]{$\rho_{22}$}
  \psfrag{p33}[][]{$\rho_{33}$}
  \includegraphics[width=0.8\hsize]{\figurelib/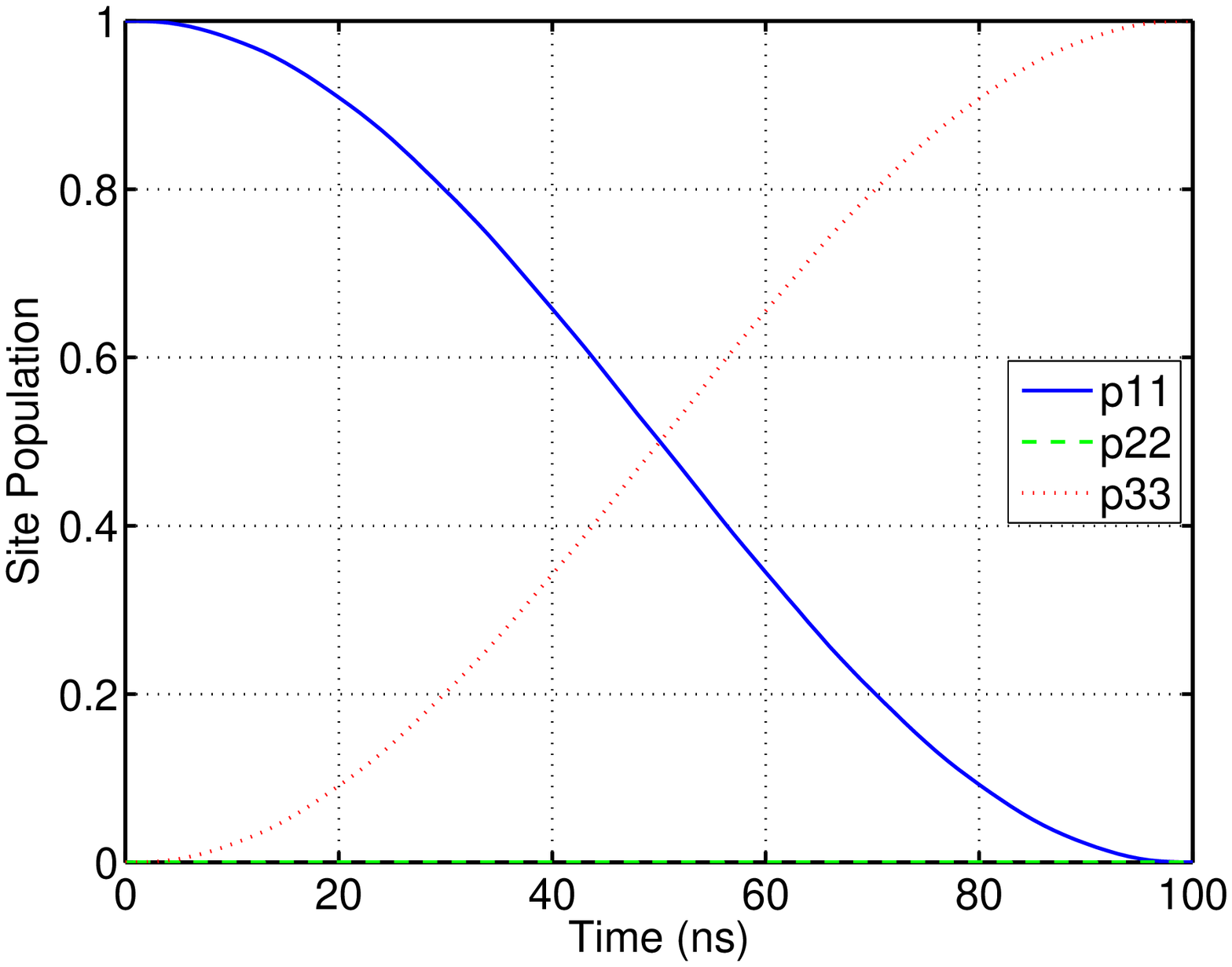}
(A)
 \psfrag{u1}[][]{$\Omega_{12}$}
 \psfrag{u2}[][]{$\Omega_{23}$}
  \includegraphics[width=0.8\hsize]{\figurelib/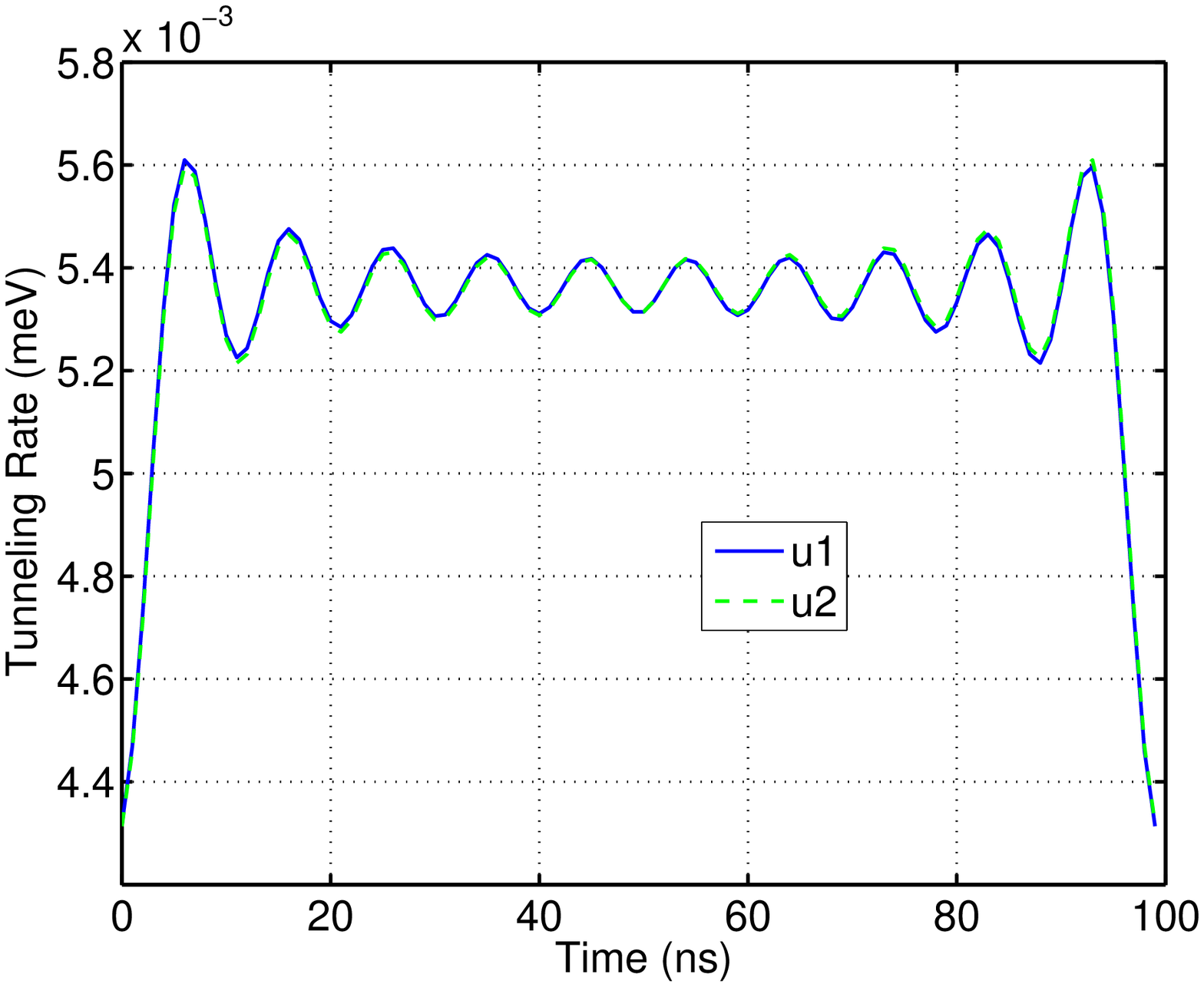}
(B)
\caption{Control pulses obtained with no robust design. 
  (A) Spatial state transfer when $\Delta=\Delta^*=2.72$ meV.  (B)
  Control pulses: blue solid line corresponds to $\Omega_{12}$, green dashed line to
  $\Omega_{23}$. (See electronic version for color plots).}
    \label{fig:1}
  \end{figure}

\begin{figure}[htb]
    \centering
 \footnotesize
\psfrag{Site Population}[][]{Site Population} 
\psfrag{Tunneling Rate (meV)}[][]{Tunneling Rate (meV)} 
 \psfrag{Time (ns)}[][]{Time (ns)} 
  \psfrag{AAA}[][]{$0.8\Delta^*$} 
  \psfrag{BBB}[][]{$1.2\Delta^*$} 
  \psfrag{p11}[][]{$\rho_{11}$} 
  \psfrag{p22}[][]{$\rho_{22}$}
  \psfrag{p33}[][]{$\rho_{33}$}
  \includegraphics[width=0.8\hsize]{\figurelib/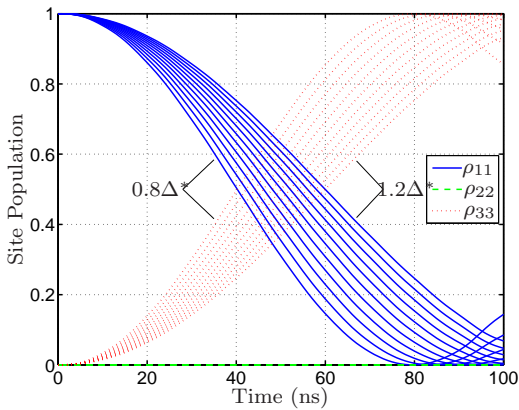}
\caption{Robustness test for the control pulses in Fig.~\ref{fig:1}:
  spatial state transfers for $11$ evenly distributed $\Delta$'s in
  the range $[0.8\Delta^*, 1.2\Delta^*]$, where $\Delta^*=2.72$ meV.}
    \label{fig:2}
  \end{figure}

In this section we apply the gradients derived in the preceding
section to design the robust control fields that can realize the
desired population transfer in solid state devices.

We consider the ionized donor chain that was discussed in
Ref.~\cite{Greentree:04}.  Typical values of the energy difference
$\Delta$ are several meV, while the control fields $\Omega_{12}$ and
$\Omega_{23}$ can be varied in the magnitude of $10^{-2}$ meV.
Realistic parameter values allow us to assume a nominal value for
$\Delta^*$ of $2.72$ meV, with the actual value of $\Delta$ deviating
from the nominal value by up to $20$\%. We further assume that the
population transfer needs to be accomplished within $100$ ns, and the
maximum feasible frequency for control fields is $0.1$ GHz. These
constraints lead to the control field expansions in \Eq{eq:10} needing
to be truncated at $M=10$.

We now discretize the total time duration $[0, 100]$ ns into $100$
small time steps, each with length $1$ ns. Take $11$ evenly
distributed sampling points from the uncertainty range $[0.8\Delta^*,
1.2\Delta^*]$ meV. Given these parameter settings, we can apply a
gradient algorithm with fixed step size to solve for the optimal
control pulses.

As a reference, we first consider the case with no robust design, \ie,
optimizing for the point $\Delta=\Delta^*$ only. The corresponding
population transfer and control fields are shown in
Fig.~\ref{fig:1}(A) and (B), respectively. To test the robustness, we
apply these control pulses to all $11$ sampling points in the
uncertainty range $[0.8\Delta^*, 1.2\Delta^*]$ meV. The results of
these simulations are plotted in Fig.~\ref{fig:2}. It is evident that
when the actual value of $\Delta$ is unknown within this range, the
electron cannot be successfully transferred from left to right, except
in the case when (coincidentally) $\Delta=\Delta^*$.  Note that in
each of the unsuccessful transfers, a full transfer is achieved at
some point before the final time $T$.  However, the oscillatory nature
of the populations leads to a reversal of the transfer.  Therefore,
experimentally, a number of different transfer times would have to be
attempted for a given pulse sequence in order to assess the
possibility of a complete transfer and to determine the optimal time.
Furthermore, noise in any element of the Hamiltonian may cause the
optimal transfer time for a given pulse sequence to be different for
each individual experiment.

\begin{figure}[htb]
    \centering
 \footnotesize
\psfrag{Site Population}[][]{Site Population} 
\psfrag{Tunneling Rate (meV)}[][]{Tunneling Rate (meV)} 
 \psfrag{Time (ns)}[][]{Time (ns)} 
  \psfrag{p11}[][]{$\rho_{11}$} 
  \psfrag{p22}[][]{$\rho_{22}$}
  \psfrag{p33}[][]{$\rho_{33}$}
  \includegraphics[width=0.8\hsize]{\figurelib/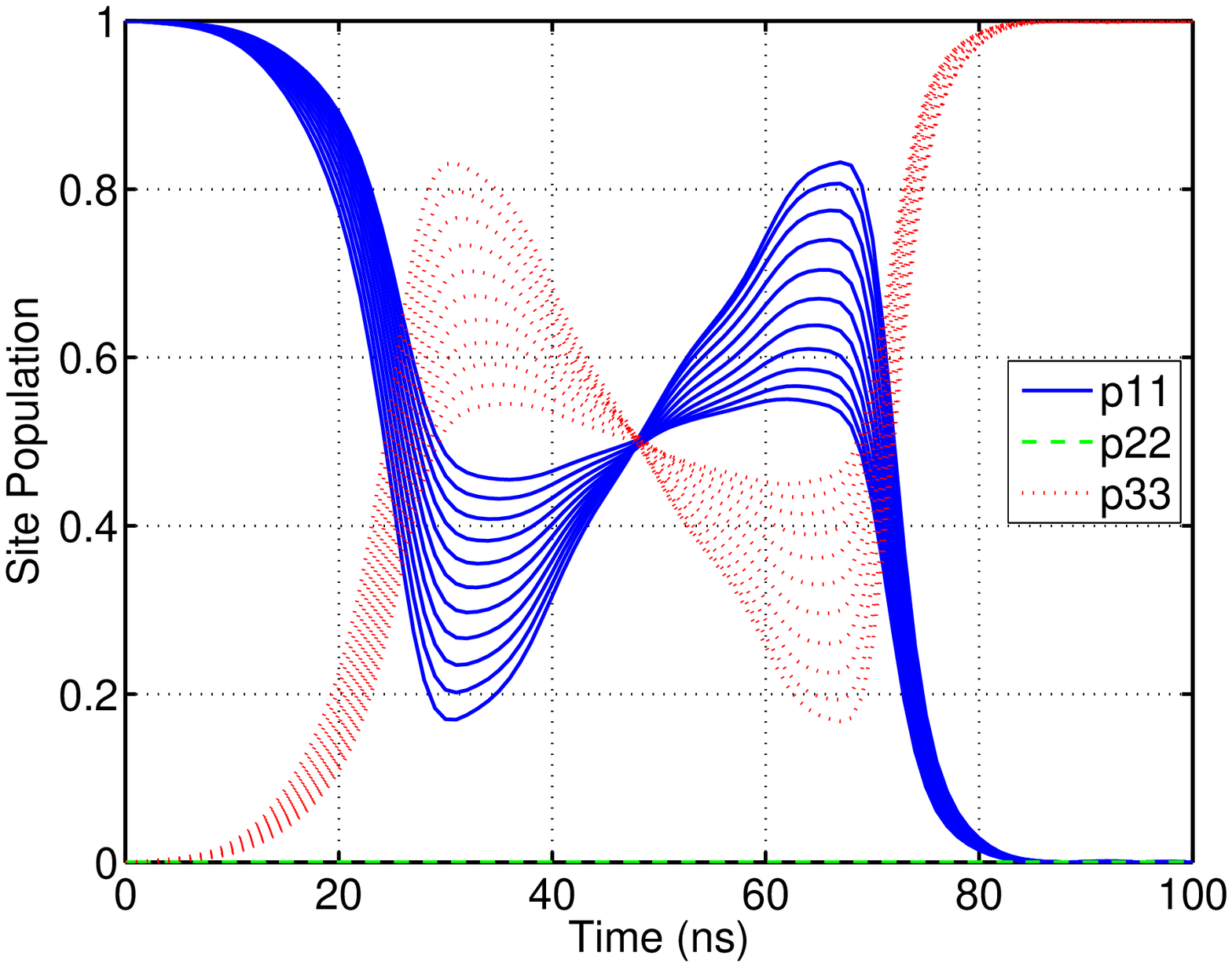}
(A)
 \psfrag{u1}[][]{$\Omega_{12}$}
 \psfrag{u2}[][]{$\Omega_{23}$}
 \psfrag{x}[][]{Time}
  \includegraphics[width=0.8\hsize]{\figurelib/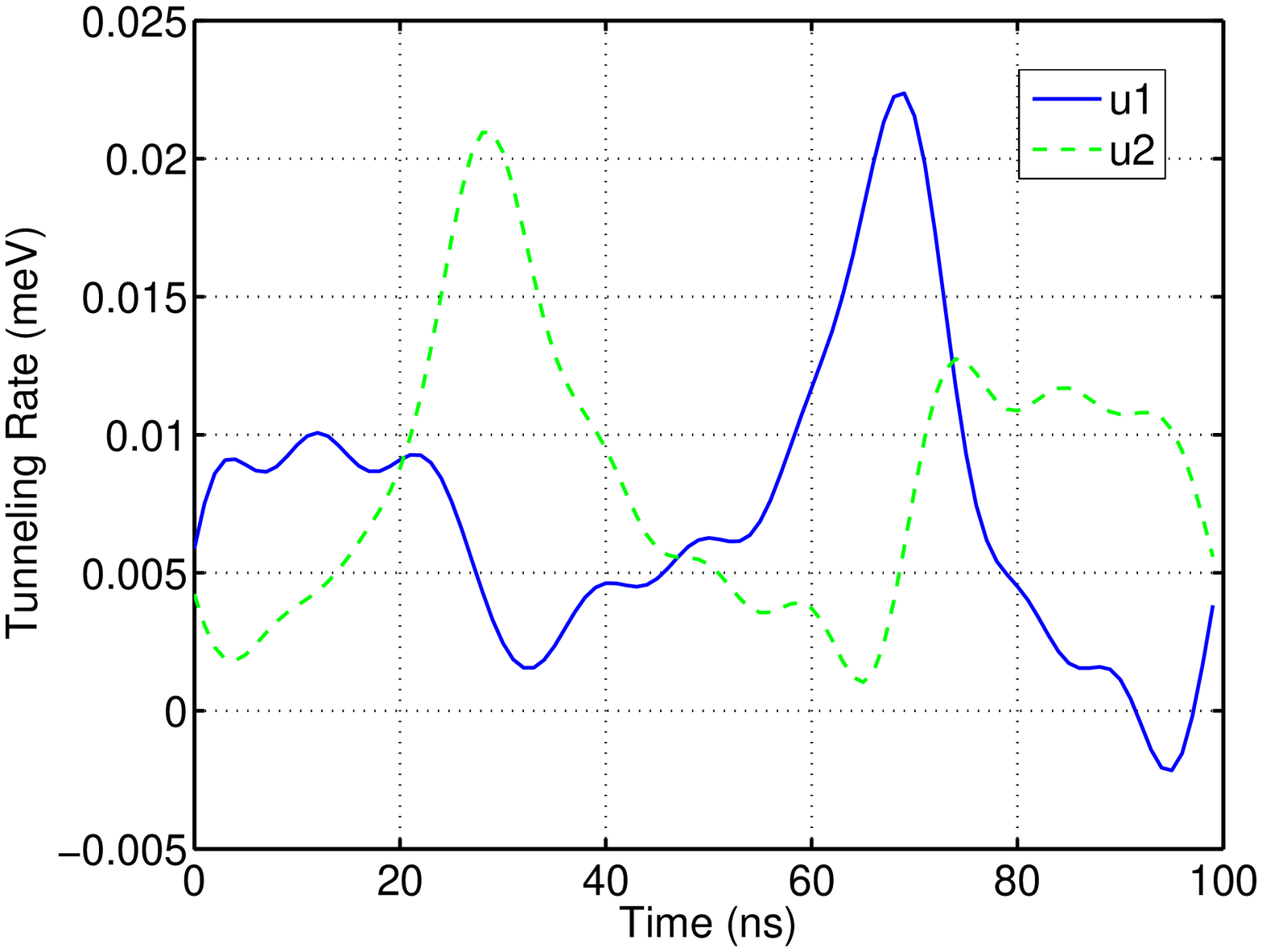}
(B)
\caption{Design of robust control fields. (A) Spatial transfer for $11$
  evenly distributed values of $\Delta$'s in the range $[0.8\Delta^*,
  1.2\Delta^*]$, where $\Delta^*=2.72$ meV.  (B) Robust control
  pulses: blue solid line corresponds to $\Omega_{12}$, green dashed line to $\Omega_{23}$. (See
  electronic version for color plots).}
    \label{fig:3}
  \end{figure}

\begin{figure}[h]
    \centering
 \footnotesize
\psfrag{Site Population}[][]{Site Population} 
\psfrag{Tunneling Rate (meV)}[][]{Tunneling Rate (meV)} 
 \psfrag{Time (ns)}[][]{Time (ns)} 
  \psfrag{p11}[][]{$\rho_{11}$} 
  \psfrag{p22}[][]{$\rho_{22}$}
  \psfrag{p33}[][]{$\rho_{33}$}
  \includegraphics[width=0.8\hsize]{\figurelib/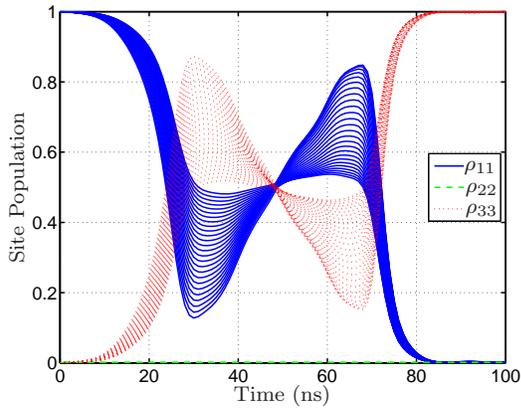}
\caption{Robustness test for $\pm 25$\% uncertainty level: spatial
  state transfers for $11$ evenly distributed $\Delta$'s in the range
  $[0.75\Delta^*, 1.25\Delta^*]$, where $\Delta^*=2.72$ meV. (See
  electronic version for color plots).}
    \label{fig:4}
  \end{figure}

  Next we apply the robust control pulses design developed above.  The
  optimization results for this scheme are shown in Fig.~\ref{fig:3}.
  The robust controls are seen to be about an order of magnitude
  larger than the controls for $\Delta^*$ only.  Most importantly, it
  is clear that whatever value of energy difference $\Delta$ within
  the $\pm 20$\% deviation range of the nominal value $\Delta^*=2.72$
  meV is employed, the resulting robust control fields can transfer
  the population with almost perfect fidelity.  The robust controls
  also have the advantage that the populations do not oscillate as in
  Fig.~\ref{fig:2}.  Slight changes in transfer time would therefore
  not affect the population transfer, a useful robustness feature from
  the experimental perspective.  Finally, we note these pulses also
  perform well outside the range for which they were defined. For
  example, if the real uncertainty level is $\pm 25$\% instead of $\pm
  20$\% in the design, the spatial state transfer still has acceptable
  performance, as shown in Fig.~\ref{fig:4}.

\section{Conclusion}
In this paper we have formulated robust control pulses designed for
electron shuttling in a chain of donors as a collection of state
transfer problems, each of which corresponds to a different value in
the uncertainty parameter range. We derived explicit formulae for the
gradients of the aggregate fidelity with respect to the control
fields, and then applied a direct gradient algorithm to solve this
problem efficiently.  The results for electron shuttling across a
three site chain show that the robust design significantly improves
the performance of an electron shuttling protocol, achieving near
perfect state transfer across a realistic range of Hamiltonian
parameters for a phosphorus-doped silicon system.

\begin{acknowledgments}
  JZ thanks the financial support from the Innovation Program of
  Shanghai Municipal Education Commission under Grant No. 11ZZ20,
  Shanghai Pujiang Program under Grant No. 11PJ1405800, NSFC under
  Grant No. 61174086, and State Key Lab of Advanced Optical
  Communication Systems and Networks, SJTU, China.  XD thanks the
  University of California-Berkeley College of Chemistry Summer
  Research Stipend. We thank NSA (Grant No. MOD713106A) for financial
  support.
\end{acknowledgments}

\bibliographystyle{apsrev}

\end{document}